\documentclass[journal,twoside,web]{IEEEtran}
\usepackage[noadjust]{cite}
\usepackage{amsmath,amssymb,amsfonts}
\usepackage{algorithmic}
\usepackage{graphicx}
\usepackage{epstopdf}
\usepackage{textcomp}
\usepackage{amsmath}
\usepackage{graphicx}
\usepackage{cite}
\usepackage{multirow}
\usepackage{booktabs} 
\usepackage{subfigure}
\usepackage{marvosym}
\usepackage{float}
\usepackage{array}
\usepackage{color}
\usepackage{url}

\begin{document}
\title{Nuclei Segmentation with Point Annotations from Pathology Images via Self-Supervised Learning and Co-Training}
\author{Yi Lin, Zhiyong Qu, Hao Chen*, \IEEEmembership{Senior Member, IEEE}, Zhongke Gao, \IEEEmembership{Senior Member, IEEE}, Yuexiang Li, \\Lili Xia, Kai Ma, Yefeng Zheng, \IEEEmembership{Fellow, IEEE}, Kwang-Ting Cheng, \IEEEmembership{Fellow, IEEE}
\thanks{The first two authors contributed equally and asterisk indicates the corresponding author.}
}

\maketitle

\begin{abstract}
Nuclei segmentation is a crucial task for whole slide image analysis in digital pathology. Generally, the segmentation performance of fully-supervised learning heavily depends on the amount and quality of the annotated data. However, it is time-consuming and expensive for professional pathologists to provide accurate pixel-level ground truth, while it is much easier to get coarse labels such as point annotations. In this paper, we propose a weakly-supervised learning method for nuclei segmentation that only requires point annotations for training. First, coarse pixel-level labels are derived from the point annotations based on the Voronoi diagram and the $k$-means clustering method to avoid overfitting. Second, a co-training strategy with an exponential moving average method is designed to refine the incomplete supervision of the coarse labels. Third, a self-supervised visual representation learning method is tailored for nuclei segmentation of pathology images that transforms the hematoxylin component images into the H\&E stained images to gain better understanding of the relationship between the nuclei and cytoplasm. We comprehensively evaluate the proposed method using two public datasets. Both visual and quantitative results demonstrate the superiority of our method to the state-of-the-art methods, and its competitive performance compared to the fully-supervised methods\footnote{Code: \url{https://github.com/hust-linyi/SC-Net}}.
\end{abstract}

\begin{IEEEkeywords}
Nuclei segmentation, Weakly-supervised, Point annotation, Self-supervised learning, Co-training.
\end{IEEEkeywords}

\section{Introduction}
\label{sec1}
\begin{figure*}[!t]
	\centering
	\subfigure[]{
		\includegraphics[width=0.12\textwidth]{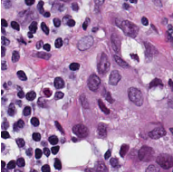}
		\label{fig_label_img}
	}
	\subfigure[]{
		\includegraphics[width=0.12\textwidth]{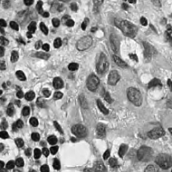}
		\label{fig_label_h}
	}
	\subfigure[]{
		\includegraphics[width=0.12\textwidth]{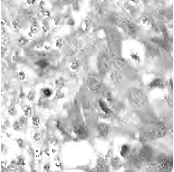}
		\label{fig_label_e}
	}
	\subfigure[]{
		\includegraphics[width=0.12\textwidth]{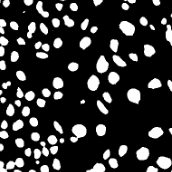}
		\label{fig_label_gt}
	}
	\subfigure[]{
		\includegraphics[width=0.12\textwidth]{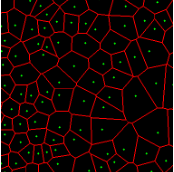}
		\label{fig_label_vor}
	}
	\subfigure[]{
		\includegraphics[width=0.12\textwidth]{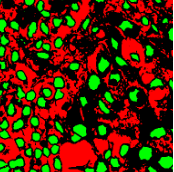}
		\label{fig_label_clu}
	}
	\subfigure[]{
		\includegraphics[width=0.12\textwidth]{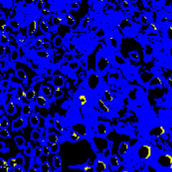}
		\label{fig_label_diff}
	}
\caption{(a) H\&E stained image; (b) H-component; (c) E-component; (d) fully annotated label; (e) Voronoi label; (f) cluster label; (g) inaccurate and incomplete area of (e) and (f), compared with (d). In (e) and (f), the green, red, and black pixels denote the label of positive, negative and ignored area, respectively. In (g), the yellow and blue masks denote the  inaccurate and incomplete area, respectively.}
	\label{fig_label}
\end{figure*}
Pathology slides contain abundant phenotypic information, and are widely used to study the manifestations of disease by analyzing cells or tissues under the microscope by pathologists~\cite{campanella2019clinical}. Nuclei segmentation is a crucial step in pathology image analysis. The shape, size, density, and other indicators of the nucleus are related to the diagnosis and treatment of cancer~\cite{irshad2013methods,graham2019hover}. Different staining methods are used to increase the contrast between the different structures for their visual examination, where the most commonly used staining method for nuclei analysis is hematoxylin-eosin (H\&E) staining, in which nuclei are stained blue-purple by hematoxylin (H-component), and cytoplasm and stromal matrix are stained red-pink by eosin (E-component)~\cite{chan2014wonderful,schmitz2021multi}, as shown in Fig.~\ref{fig_label_img}. Besides, with the advent of dedicated scanners, slices can be easily converted to digital pathology images, which are convenient to store on computers for further processing and analysis. Therefore, the wide availability of digital H\&E stained pathology images greatly facilitates researchers in developing and validating advanced automatic pathological image analysis methods with nuclei segmentation.

Traditional nuclei segmentation algorithms include water-shed~\cite{shu2013segmenting}, thresholding~\cite{win2017automated},  and other morphological operations~\cite{yin2017tumor}, which usually require a series of pre-processing methods such as defuzziﬁcation and contrast enhancement to improve the image quality. 
In recent years, methods based on deep learning have made significant progress~\cite{long2015fully,ronneberger2015u,zhou2019cia,che2023iqad}. 
A number of algorithms based on models such as fully convolutional networks (FCN)~\cite{long2015fully} and U-Net~\cite{ronneberger2015u} have been applied to medical image segmentation tasks. 
Especially U-Net and its variants have proven their effectiveness in nuclei segmentation~\cite{kumar2017dataset,raza2019micro,lin2022insmix,koohbanani2020nuclick}. In addition, the contour-based method has been used to predict more accurate nucleus boundaries~\cite{zhou2019cia,chen2017dcan,graham2019hover,raza2019micro,naylor2018segmentation}, which helps split the touched and overlapped nuclei.

Although the deep learning based segmentation methods have demonstrated promising performance, the fully-supervised training of deep learning methods requires a large amount of data with pixel-level annotations. Obtaining such ground truth data is challenging as pixel-accurate annotation is very time-consuming and requires professional clinical knowledge. The limited availability of data with pixel-level annotation thus makes nuclei segmentation still a challenge. To conquer this problem, weakly-supervised methods have been widely investigated, which significantly reduce the burden of manual annotation, as they only rely on weak labels in different granularities. Specifically, polygon labels were used in~\cite{zhao2018deep} where the object detection response is propagated to the segmentation mask. Scribble labels were used in~\cite{lee2020scribble2label} that utilized a self-training strategy to propagate the weak labels. Some studies ~\cite{Dong2019ICCV,feng2017discriminative} attempted to use image-level labels only for segmentation. However, due to the small size and a large number of nuclei, these commonly used weak labels are not effective for the task of nuclei segmentation. For example, the polygon and scribble labels still require cumbersome annotation labor, and the image tags cannot provide location information for the enormous quantity of nuclei.

Point annotation can be viewed as the most efficient manner for annotation, gaining increasing attention in cell/nuclei segmentation. Existing methods~\cite{chamanzar2020weakly,tian2020weakly,qu2019weakly,yoo2019pseudoedgenet} mainly encode the point annotation into coarse pixel-level labels, such as point distance maps~\cite{chamanzar2020weakly}, Voronoi labels~\cite{tian2020weakly}, cluster labels~\cite{qu2019weakly}, and pseudo edge maps~\cite{yoo2019pseudoedgenet}. As shown in Fig.~\ref{fig_label}, these methods typically inject the shape and texture prior knowledge of nuclei into coarse pixel-level label. For example, the points are assumed to be around the center and the nuclei shapes are nearly convex. Despite drastically reducing the annotation cost and alleviating the data-imbalance problem of point annotation, these methods still suffer from incomplete coarse labels and inaccurate boundary information, as illustrated in Fig.~\ref{fig_label_diff}. Various methods have been proposed to eliminate the distraction induced by the coarse labels, which can be categorized into two classes: 1) multi-stage optimization to refine the segmentation in a bootstrapping way~\cite{chamanzar2020weakly,Qu2020,tian2020weakly}; 2) adding additional constraints based on the local contrast between nuclei and their surrounding cytoplasm~\cite{qu2019weakly,yoo2019pseudoedgenet}. However, the additional constraints could introduce inaccurate supervision, and a multi-stage learning strategy may suffer from error accumulation (i.e., the global optimum cannot be guaranteed).

To overcome the challenge of multi-stage optimization, one potential end-to-end training solution is collaborative training (co-training). As a common method in semi-supervised learning that jointly uses labeled and unlabeled data to improve the generalization of the model through collaboration among multiple learners~\cite{Ning2021}, it has been broadly applied to image classiﬁcation~\cite{hong2015spatial}, target recognition~\cite{qiao2018deep} and image segmentation~\cite{wang2021self}. For medical image analysis, co-training has been adopted for semi-supervised leaning~\cite{Huang2020}
or consistency learning in different views~\cite{qiao2018deep}. For cell segmentation, \cite{zhao2020institute} combined co-training with divergence loss to enlarge the prediction difference between the two models. They further introduced a divergence loss to avoid the self-deception problem, i.e., the two networks overfitting to each other. However, the divergence loss is in contrast to the fundamental loss and may hurt the performance of the individual model. Moreover, when the cell area has a low contrast or the cell shape is irregular, it would be difficult to obtain a precise boundary. To address this challenge, in this paper, we introduce an exponential moving average (EMA) to periodically average pseudo labels in the co-training process, and further combine them with coarse labels to achieve a more stabilized and accurate cross-supervision. 
\begin{figure*}[!t]
	\centering
	\includegraphics[width=0.98\textwidth]{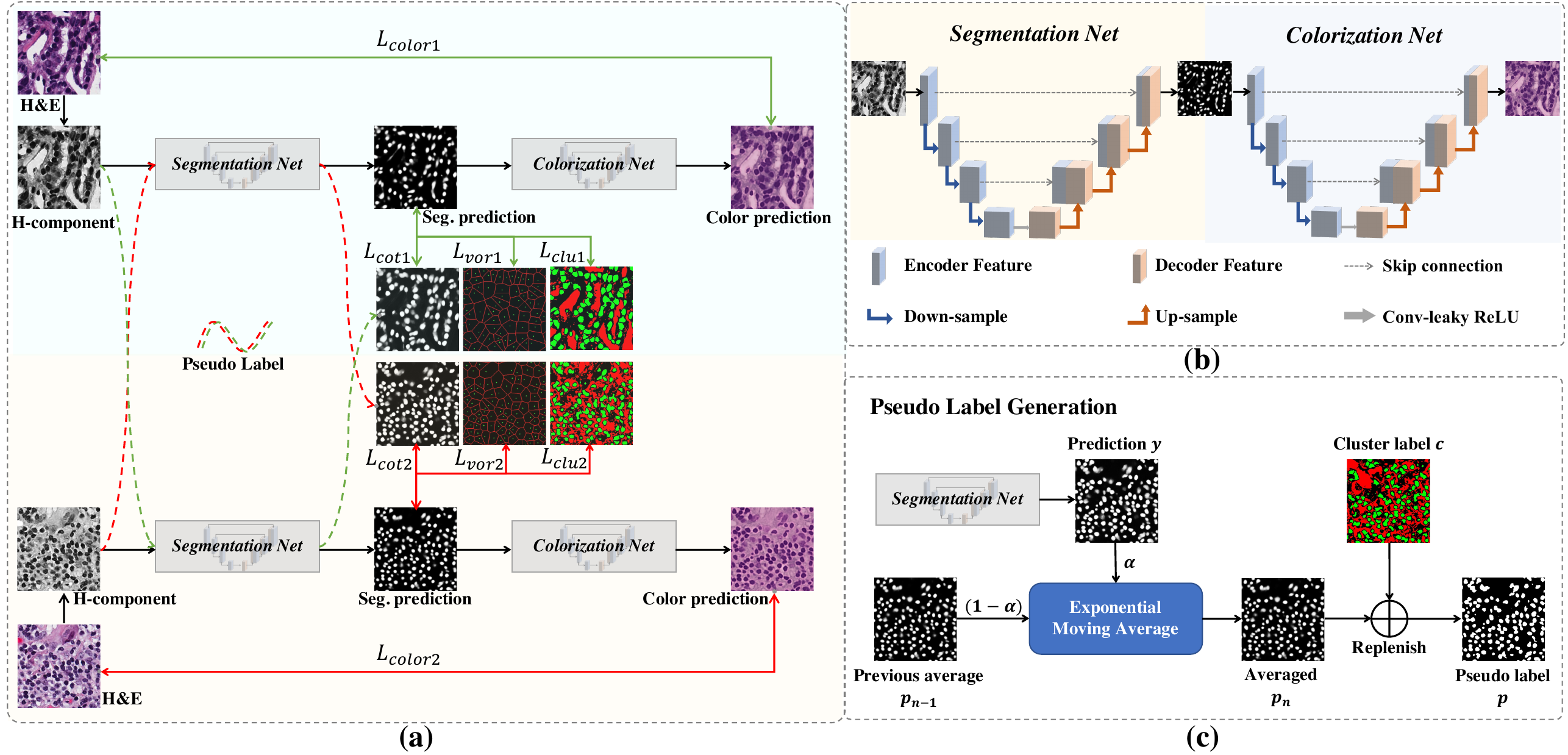}
	\caption{The framework of the proposed method.
		(a) The pipeline of the proposed method; (b) The framework of SC-Net; (c) The process of pseudo label generation. $L_{vor}$, $L_{clu}$, $L_{cot}$, $L_{color}$ denote the Voronoi loss, cluster loss, co-training loss, and colorization loss, respectively.}
	\label{fig_framework}
\end{figure*}

Addressing the challenge caused by imposing additional constraints, instead of directly treating the additional supervision as a golden standard, self-supervised learning (SSL) provides a more effective way to exploit the implicit supervision. Recently, we have witnessed the advances of SSL in natural image analysis\cite{chen2020simple,he2020momentum,che2023DGDR,grill2020bootstrap}, which can be categorized into three types: contrastive learning~\cite{chen2020simple,he2020momentum}, clustering~\cite{Huang2020}, and consistency learning~\cite{grill2020bootstrap}. Extensive recent  studies~\cite{koohbanani2021self,li2021dual,srinidhi2022self} in pathology image analysis have demonstrated the effectiveness of SSL techniques, which were originally proposed for natural images (e.g., CPC~\cite{oord2018representation}, SimCLR~\cite{chen2020simple}, and MoCo~\cite{he2020momentum}). Subsequently, instead of straightforward extensions, recent studies attempted to develop SSL methods addressing unique problems encountered in pathology images, such as gigapixel whole slide image (WSI)~\cite{li2021dual} and nuclei counting~\cite{xie2020instance}. However, these contrastive-learning-based methods could suffer from the opposite semantic labels~\cite{li2021sslp} if the anchor is located on the boundary of target tissues or cells, as well as the severe data imbalance problem between the positive and negative pairs. To address these problems, this paper proposes a generative SSL method for nuclei segmentation of the H\&E stained pathology images. This custom-made method benefites the nuclei segmentation network with an auxiliary network, i.e., transforming the H-components into the H\&E images, which encourages the model to exploit the relationship between nuclei and surrounding cytoplasm. In the previous SSL methods~\cite{yang2021self,xie2020instance,li2021sslp}, the network is pre-trained with the pretext task, and then fine-tuned for the target task. The discrepancy between the pretext task and the target task will reduce the effect of self-supervision learning during model transferring~\cite{he2019rethinking,koohbanani2021self}. Furthermore, the same network has to be used for both tasks~\cite{che2022learning}. The proposed self-supervised setting jointly train the pretext task and the target task in an end-to-end manner, reducing the impact of task discrepancy.

In this paper, we aim at an end-to-end weakly-supervised nuclei segmentation framework based only on point annotations, as illustrated in Fig.~\ref{fig_framework}. The primary idea behind the method is label propagation where the original point labels are propagated to the pixel-level labels in a coarse-to-fine manner. Specifically, the proposed method consists of three parts as follows. 
First, we convert the point annotations into coarse pixel-level labels by training the initial segmentation network based on the Voronoi diagram and the distance-aware $k$-means clustering algorithm. The coarse pixel-level labels have complementary information with respect to the point labels, while also containing errors and uncertainties.
Second, to minimize the distraction caused by the incomplete coarse labels, we design a co-training method, in which a pair of networks supervise each other with the pseudo labels. In this way, the two networks can mutually transfer knowledge to each other and boost the performance. We periodically average the pseudo labels in the training process using EMA to achieve a more stabilized and accurate cross-supervision. 
Third, to further explore the boundary information which is omitted by the inaccurate coarse labels, we propose a self-supervised representation learning method that transforms H-components into the H\&E images. Based on the fact that eﬀective representations are restored in the stain-separated H-components, this nuclei-aware colorization proxy task can extract more details, including the correlation between nuclei and cytoplasm. To this end, a novel end-to-end self-supervised framework is designed, which sequentially combines a \textbf{S}egmentation network with a \textbf{C}olorization network, named SC-Net. The colorization network helps the segmentation network learn to become implicitly self-aware of the nuclei boundaries from the transformation from H-components into H\&E images without a need for manual annotations. 

In addition, to balance the learning focus in the training process, we introduce a cumulative learning strategy to integrate the whole pipeline, which
is designed based on two key principles: 1) the learning focus of SC-Net should gradually change from the representation learning (i.e., colorization) to the target task (i.e., segmentation); and 2) with the improvement of the segmentation network, the corresponding pseudo labels become more trustworthy and should thus be assigned a greater weight during training. 

The proposed method is validated on two public datasets, the Multi-Organ Nuclei Segmentation (MoNuSeg) dataset~\cite{kumar2017dataset,kumar2019multi} and the Computational Precision Medicine (CPM) dataset~\cite{vu2019methods}. Extensive experiments consistently demonstrate the effectiveness of each module and the superiority of our model to the state-of-the-art methods. In summary, the major contributions of this work are three-fold:
\begin{itemize}
    \item \textcolor{black}{We propose a novel weakly-supervised nuclei segmentation framework using only point annotations. The proposed method propagates the point labels into pixel-level labels in a coarse-to-fine manner.}
    \item We design a co-training strategy, where a pseudo label generation module, leveraging  the EMA method to stabilize the pseudo labels, successfully supplies incomplete coarse labels.
    \item We propose a self-supervised method tailored for nuclei segmentation in pathology images, with a novel SC-Net to be self-aware of the relationship between nuclei and cytoplasm.
\end{itemize}

\section{METHODOLOGY}
In this section, we describe the proposed weakly-supervised nuclei segmentation method using point labels with label propagation,
as shown in Fig.~\ref{fig_framework}(a), which consists of three major modules: 1) nuclei segmentation based on coarse pixel-level labels generated from the point labels; 2) a co-training strategy based on exponential moving average (EMA) for generating the pseudo labels, and 3) self-supervised representation learning with H-component based colorization. In the following, we will detail each module.

\subsection{Nuclei Segmentation with Coarse Labels}
\subsubsection{Coarse pixel-label generation from point label}
\textcolor{black}{In practice, directly using point annotation for the nuclei segmentation could suffer from the data-imbalance problem due to insufficient supervision information. To alleviate this issue, we adopt the Voronoi diagram~\cite{tian2020weakly} and the $k$-means clustering method~\cite{qu2019weakly} to generate Voronoi label and cluster label, respectively. Specifically, for Voronoi labels, as shown in Fig.~\ref{fig_label_vor}, the image is divided into convex polygons based on the point annotations with the assumptions that the point labels are at the nuclei centers and nuclear shapes are convex (even though these assumptions are not necessarily valid). For cluster labels, as shown in Fig.~\ref{fig_label_clu}, we utilize the $k$-means to obtain more supervision information of nuclei boundary and shape as follows. First, we calculate the distance maps from the point labels by performing distance transform between each pair of points. Second, the distance maps are combined with the original H\&E stained images to conduct the $k$-means clustering to divide all pixels into $k=3$ clusters: nuclei, background, and ignored area.
The clusters that have maximum and minimum overlap with the point annotations are labeled as nuclei and background, respectively. 
The remaining one is the ignored class. 
The introduction of an ignored area allows pixels that cannot be easily determined as nuclei or background with certainty not be forced into either class, ensuring that the clustering can assign correct pixel labels as much as possible~\cite{Qu2020}. Third, several morphological operations are adopted to refine the cluster label, including connected domain labeling, scattered region removing, morphological opening operation, and binary hole filling.}

\subsubsection{Hematoxylin Component Extraction}
According to the principle of H\&E staining~\cite{chan2014wonderful}, H-component of the original pathology image can provide sufficient information for nuclei segmentation and reduce the stain variance as well. Thus, we apply the stain separation method~\cite{vahadane2016structure} to separate H-components and E-components from the original color images. 

Specifically, stain separation is an estimation of the density map from the colors at each pixel. The stain illumination intensity in a certain spectrum depends on the tissue type and amount of stain absorbed. This relationship can be captured in the Beer-Lambert law~\cite{loos2008multiple} as follows:
\begin{equation}
    x = x_0\exp(-WD),
\end{equation}
where $x$ and $x_0$ denote the matrix of colors (i.e., three channels corresponding to red, green, and blue) and illumination, respectively, $W$ denotes the stain color appearance matrix which represents the color basis of each stain, and $D$ denotes the stain density map which represents the concentration of each stain. And we utilize the non-negative matrix factorization method~\cite{newberg2008framework} to estimate $D$ and $W$. Then, the H-component $x_h$ and the E-component $x_e$ can be extracted from the stain color appearance matrix $D$ as follows:
\begin{equation}
    x_h = x_0\exp(-D[0, :]),\; x_e = x_0\exp(-D[1, :]).
\end{equation}

As shown in Fig.~\ref{fig_label_h}, although part of the color information is lost in the obtained H-component image, the color contrast between the nuclear and non-nuclear regions is enhanced. 

\subsubsection{Nuclei Segmentation with Coarse Labels}
\textcolor{black}{Using the derived pixel-level coarse labels (i.e., the Voronoi and cluster labels) could potentially achieve reasonable results for nuclei segmentation. In this paper, we adopt the ResUNet~\cite{Qu2020} as the segmentation network which integrates the residual blocks~\cite{he2016deep} into the U-Net~\cite{ronneberger2015u}, as shown in Fig.~\ref{fig_framework}(b). The segmentation network is trained with cross-entropy loss with respect to the Voronoi label and the cluster label:}
\begin{equation}
L_{vor} = - \frac{1}{{|\Omega_v|}}\sum\limits_{i \in \Omega_v} \left[v_i \cdot \log y_i + (1 - v_i) \cdot \log (1 - y_i) \right],
\label{eq_vor}
\end{equation}
\begin{equation}
L_{clu} = - \frac{1}{{|\Omega_c|}}\sum\limits_{i \in \Omega_c} \left[c_i \cdot \log y_i + (1 - c_i) \cdot \log (1 - y_i) \right],
\label{eq_clu}
\end{equation}
where $v_i$ and $c_i$ denote the  Voronoi label and cluster label of nuclei at the $i$-th pixel, respectively; $y=S(x_h)$ denotes the prediction of the segmentation network $S$ with H-component $x_h$ as input; and $\Omega_*$ (i.e., $\Omega_v$ or $\Omega_c$) is the set of non-ignored pixels. As illustrated in Fig.~\ref{fig_label}, the Voronoi label is used to supervise the network to separate overlapping nuclei, while the cluster label could provide coarse shape and boundary information for nuclei segmentation. In the following, we elaborate the co-training strategy to further provide supervision to the ignored area of the coarse cluster labels, denoted with the blue color in Fig.~\ref{fig_label_diff}.

\subsection{Co-training for Incomplete Coarse Cluster Labels}
With more ``annotated" pixels in the cluster label, the network can learn some nuclei regions near the point annotation better, but it also can be misled by the error of the cluster label. To address this challenge, we introduce a co-training strategy to minimize the distraction of the erroneous cluster label. 

The co-training framework consists of a pair of segmentation networks $S_a$ and $S_b$, which are trained by two subsets of training data, i.e., $X_a$ and $X_b$, respectively. To encourage the two networks to learn different and supplementary information from the data, 
we divide the training set $X$ into two non-overlapping subsets of equal size, i.e., ${X_a} \cup {X_b} = X$ and ${X_a} \cap {X_b} = \emptyset$. 
In the following, we  detail the training procedure of the first segmentation network $S_a$ supervised by $S_b$. The training of the second segmentation network $S_b$ supervised by $S_a$ follows a similar procedure.
In the training process of $S_a$ with the subset $X_a$, besides the aforementioned coarse labels (i.e., Voronoi and cluster), we further utilize the pseudo labels generated by $S_b$.
To stabilize the co-training process, instead of directly using the prediction of another network as the pseudo label~\cite{zhao2019weakly}, we design a pseudo label generation strategy as follows. First, we periodically calculate EMA of the predictions to obtain more robust pseudo labels:  
\begin{equation}
    p_n = \alpha y_b + (1-\alpha)p_{n-1},
\end{equation}
\textcolor{black}{where $\alpha$ is the EMA weight (which is empirically set to 0.1), $y_b=S_b((x_h)_a)$ is the prediction of $S_b$ with respect to the H-component $(x_h)_a$, and $n$ denotes the time step for the moving average, which means how many steps the predictions are averaged. To reduce the computational cost, we average the predictions every $\gamma$ epochs (which is empirically set to 3). Then, the averaged predictions are used to determine the labels for the ignored area of the cluster label:}
\begin{equation}
p_i = \left\{ {\begin{array}{*{20}{c}}
	{{c_i}}&{\mathrm{if} \ {c_i} = 0 \ \mathrm{or}\ {c_i} = 1}\\
	{p_i}&{c_i=2}
	\end{array}} \right.
\label{...}
\end{equation}
\textcolor{black}{where $c_i=[0,1,2]$ denotes the cluster label of the $i$-th pixel, and the left-hand $p_i$ is the combination of the pseudo label and cluster label. The cluster label of 0, 1, 2 means background, foreground, and ignored area, respectively. We calculate the co-training loss with the Kullback–Leibler (KL) divergence:}
\begin{equation}
\label{eq_cot}
    L_{cot1} = \frac{1}{|\Omega|} \sum\limits_{i \in \Omega} p_i \cdot \log \frac{p_i}{(y_a)_i},
\end{equation}
where $p_i$ is the $i$-th pixel of the pseudo label, $\Omega$ is the set of all pixels, and $y_a=S_a((x_h)_a)$ denotes the predictions of $S_a$ with respect to the H-component $(x_h)_a$. With the training set $X$ split into two non-overlapping subsets $X_a$ and $X_b$, two segmentation networks are trained with the coarse labels (i.e., the Voronoi and cluster labels) by $X_a$ and $X_b$, respectively, the two networks could transfer knowledge to each other with their respective pseudo labels, so as to compensate for the missing supervision information. However, the segmentation performance may still suffer from the inaccurate cluster label, as shown in Fig.~\ref{fig_label_diff}. Hence, our next step is to explore an auxiliary colorization task that transforms H-component images back into the original H\&E stained images from which more precise nuclei boundaries can be obtained.

\subsection{Self-supervised Nuclei-aware Colorization}
To exploit nuclei boundary for segmentation network, we design a self-supervised visual representation learning method tailored for nuclei segmentation in pathology images by image colorization based on the H-component. The proposed pipeline consists of two U-Nets in a sequential order, as shown in Fig.~\ref{fig_framework}(b), which first generates the probability map of the nuclei from the H-component image, followed by a colorization network that reconstructs the original H\&E image from the probability map. Note that the segmentation procedure can be explicitly trained by the Voronoi label $L_{vor}$ and the cluster label $L_{clu}$ in Eq.~(\ref{eq_vor}), as well as the co-training loss in Eq.~(\ref{eq_cot}), while the colorization procedure can be trained in an unsupervised manner that converts the H-components to the H\&E stained images. Besides, the segmentation network and the colorization network are connected by the probability map of the nuclei, which achieves an end-to-end training and the colorization task would implicitly promote the nuclei representation learning of the segmentation network. The colorization loss is computed between the predicted and true images as:
\begin{equation}
    \label{eq_color}
    L_{color} = \frac{1}{|\Omega|} \sum\limits_{i \in \Omega} \left\| C\left(S\left(x_h\right)\right)_i - x_i\right\|^2_2,
\end{equation}
where $x_i$ denotes the $i$-th pixel of the original H\&E stained image, $\Omega$ denotes the set of all pixels, $C(\cdot)$ denotes the prediction of the colorization network, and $S(x_h)$ denotes the probability map generated by the segmentation network from the H-component image $x_h$ extracted from $x$. By solving this proposed colorization pretext task, the segmentation network can capture more low-level features from the H-component image, and model the relationship between the nuclei and cytoplasm in an implicit way. 

\subsection{Integration}
The proposed method integrates the knowledge learned by the different modules under a cumulative learning framework. Specifically, adaptive trade-off parameters (i.e., $\alpha$ and $\beta$) are adopted to control the weights of different losses. The final loss can be formulated as:
\begin{equation}
    \label{eq_final}
    \begin{aligned}
        L_{total} = &{L_{SC1}} + {L_{SC2}} \\
        = &L_{vor1} + L_{clu1} + \alpha L_{cot1} + \beta L_{color1} + \\
        &L_{vor2} + L_{clu2} + \alpha L_{cot2} + \beta L_{color2},
    \end{aligned}
\end{equation}
where ${L_{SC1}}$ and ${L_{SC2}}$ denote the loss of each SC-Net in the co-training procedure. Considering that the quality of pseudo labels will be gradually improved during training, the weight of the co-training loss should be increased accordingly. Hence, following~\cite{zhou2020bbn}, the trade-off parameter $\alpha$ is set as $\alpha = \eta \left(\frac{N}{N_{\max}}\right)^2$,
where $N$ and $N_{\max}$ are the current epoch and the total number of training epochs, respectively, and $\eta$ is a constant.  On the other hand, the training focus should gradually change from the representation learning (colorization) to the target task (segmentation), the trade-off parameter $\beta$ should gradually decrease to facilitate the segmentation task as $\beta = \epsilon \left(1 - \frac{N}{N_{\max}}\right)^2$,
where $\epsilon$ is a constant.

\section{Experiments and Results}
\subsection{Datasets}
We evaluate our proposed method on the Multi-Organ Nuclei Segmentation (MoNuSeg) dataset and the Computational Precision Medicine (CPM) dataset. Both datasets have pixel-level annotation of the nuclei, and we obtained the point annotation by extracting the point roughly around the center of each mask as~\cite{tian2020weakly,qu2019weakly,chamanzar2020weakly,Qu2020}.

\textbf{MoNuSeg} is a public dataset obtained by annotating the pathology images for tumors in different organs, which consists of H\&E stained images captured at 40x magnification downloaded from the cancer genome atlas (TCGA).\footnote{https://www.cancer.gov/tcga} It contains 30 training images and 14 test images, each with the size of $1000\times1000$ pixels. We randomly select 6 images in the training set for validation. The training data contains approximately 22,000 nuclei with boundary fully annotated, and the test data contains around 7,000 annotated nuclei. 

\textbf{CPM} contains 32 images with the size of $500\times500$ or $600\times600$ pixels, scanned at 40$\times$ magnification. It contains four types of tumor, i.e., non-small cell lung cancer, head and neck squamous cell carcinoma, glioblastoma multiforme, and lower grade glioma. For each type of tumor, we randomly select one image for validation and two images for testing, thus all the images are divided into 20 images for training, 4 images for validation, and 8 images for testing.

\subsection{Implementation Details and Evaluation Metrics}
For data preprocessing, we crop the training images to patches of $250 \times 250$ pixels with an overlap of 125 pixels and the testing images to patches of $224 \times 224$ pixels with an overlap of 80 pixels. All the training patches are then randomly cropped to $224 \times 224$ pixels, zoomed, flipped, and rotated before feeding to the models.
The backbone used is ResNet-34~\cite{he2016deep} pre-trained on ImageNet dataset~\cite{russakovsky2015imagenet}.
We train the network with the learning rate reduced by a factor 10 every 30 epochs with an initial value of $1\times10^{-3}$. The Adam optimizer is adopted, with weight decay $5\times10^{-4}$.
The loss function weights $\eta$ and $\epsilon$ are empirically set as 1 and 0.1, respectively. 
During the inference stage, we only use the segmentation network producing the segmentation probabilities, and discard the colorization network. The two trained models with the lowest loss on validation set are integrated to give an output. 

\textcolor{black}{
Following~\cite{Qu2020}, We select four commonly used metrics to evaluate our method, including two pixel-level metrics (i.e., pixel accuracy and F1 score) and two object-level metrics (i.e., objective-level Dice coefficient ($\mathrm{Dice}_{obj}$)~\cite{Sirinukunwattana2015}, aggregated Jaccard index (AJI)~\cite{kumar2017dataset}). It is worth noting that $\mathrm{Dice}_{obj}$ is often used to evaluate the overlapping between the prediction and ground-truth, and AJI is a widely accepted method to quantify the instance-level segmentation performance~\cite{kumar2019multi,huang20212}.
Following~\cite{kirillov2019panoptic,graham2019hover}, we also adopt three evaluation metrics to evaluate our approach and other methods, including Detection Quality (DQ), Segmentation Quality (SQ), and Panoptic Quality (PQ).
}
\subsection{Results and Comparison}
\subsubsection{Comparison with Weakly-Supervised Methods}

\begin{table*}[thbp]
	\centering
	\color{black}
	\caption{Comparison with other weakly-supervised methods with point annotation.}
        \renewcommand\arraystretch{1.2}
	\setlength{\tabcolsep}{6pt}{
		\begin{tabular}{l|ccccccc|ccccccc}
			\toprule
			 \multirow{3}{*}{Model} & \multicolumn{7}{c|}{MoNuSeg}	& \multicolumn{7}{c}{CPM} \\
			 \cline{2-15}  & Acc   & F1   & $\mathrm{Dice}_{obj}$  & AJI  & DQ  & SQ  & PQ  & Acc   & F1   & $\mathrm{Dice}_{obj}$  & AJI  & DQ  & SQ  & PQ  \\
			   & (\%) & (\%) & (\%) & (\%) & (\%) & (\%) & (\%) & (\%) & (\%) & (\%) & (\%) & (\%) & (\%) & (\%) \\
			 \hline
			 \hline
			Yoo~\textit{te al.}~\cite{yoo2019pseudoedgenet}        & 86.51	& 72.11	& 56.68	& 29.03 & 43.97    & 72.21 & 31.86 & 90.50 & 79.81    & 72.46  & 49.42 & 61.04	& \textbf{73.63}	& 45.10\\
			Tian~\textit{te al.}~\cite{tian2020weakly}              & 88.01  & 71.47 & 63.96 & 40.51 & 50.67 & 67.19 & 34.12 & 87.87 & 71.74 & 64.39 & 42.11 & 42.86	& 63.03	& 27.06 \\ 
		    Xie~\textit{te al.}~\cite{xie2020instance}             & 91.19	& 77.56	& 72.51	& 51.69 & 68.82	& 72.50	& 49.97 & 89.96 & 75.87 & 70.28 & 49.40 & 59.49 & 70.87 & 42.68 \\
			Qu~\textit{te al.}~\cite{Qu2020}                      & \textbf{91.52}	& 76.76	& 73.24	& 54.32 & 69.72	& 71.28	& 49.84 & 89.90 & 76.56 & 71.17 & 50.91 & 64.10	& 70.66	& 45.69 \\
	    	Cha.~\textit{te al.}~\cite{chamanzar2020weakly}         & 91.04	& 74.18	& 71.70	& 53.69 & 69.40	& 69.84	& 48.75 & 88.57 & 70.69 & 66.44 & 45.92 & 57.06	& 67.80	& 39.20 \\
			Lee~\textit{te al.}~\cite{lee2020scribble2label}       & 91.13	& 77.05	& 73.44	& 54.20 & 72.03	& 71.80	& 51.78 & 89.85 & 75.80 & 70.82 & 50.26 & 65.37	& 69.75	& 45.99 \\
			\hline
			\textbf{Ours}    & 91.44    & \textbf{77.64}    & \textbf{74.41}    & \textbf{56.20} &\textbf{73.27}  	&\textbf{72.48}  	& \textbf{53.19}  &\textbf{91.01} 	& \textbf{79.97} & \textbf{73.73} &\textbf{51.69} & \textbf{68.42} & 72.18 &  \textbf{49.66}\\
			\bottomrule
		\end{tabular}}%
	\label{tab_sota}%
\end{table*}%

\begin{table}[thbp]
	\centering
	\color{black}
	\caption{The segmentation results of different Organ on MoNuSeg.}
	\resizebox{0.48\textwidth}{!}{
		\begin{tabular}{lccccccc}
			\toprule
			 \multirow{2}{*}{Organ}  & Acc   & F1   & $\mathrm{Dice}_{obj}$  & AJI & DQ  & SQ  & PQ \\
			  & (\%) & (\%) & (\%) & (\%) & (\%) & (\%) & (\%) \\
			 \midrule
			    Bladder & 90.74 & 79.99 & 76.95 & 58.76 & 74.19 & 75.49 & 56.00 \\
                Brain & 94.19 & 80.10 & 78.01 & 61.63 & 81.87 & 71.61 & 58.59 \\
                Breast & 88.01 & 73.34 & 69.20 & 49.33 & 64.69 & 67.50 & 43.66 \\
                Colon & 93.17 & 75.60 & 73.88 & 56.51 & 72.81 & 71.48 & 52.05 \\
                Kidney & 91.72 & 78.01 & 75.55 & 57.91 & 76.80 & 73.38 & 56.40 \\
                Lung & 90.05 & 76.69 & 72.53 & 54.84 & 71.82 & 73.36 & 52.83 \\
                Prostate & 93.76 & 79.08 & 75.76 & 57.04 & 74.46 & 74.34 & 55.37 \\
			\bottomrule
		\end{tabular}}%
	\label{tab_organ}%
\end{table}%

\textcolor{black}{
We compare our method with several representative weakly-supervised nuclei segmentation methods. 
Table~\ref{tab_sota} presents the segmentation results of several state-of-the-art weakly-supervised methods~\cite{lee2020scribble2label,Qu2020,chamanzar2020weakly,tian2020weakly,yoo2019pseudoedgenet,xie2020instance}, of which ~\cite{lee2020scribble2label,xie2020instance} have no released code, and we reimplement the methods with the same backbone and hyper-parameters for a fair comparison.
It is obvious from Table~\ref{tab_sota} that our method achieves superior segmentation performance compared with other weakly-supervised methods, and the improvements to $\mathrm{Dice}_{obj}$, AJI , and PQ are significant. 
In Table~\ref{tab_organ}, we provide the organ-wise segmentation results of the proposed methods. It can be seen that the performance on breast images is inferior to other organs.
}

\textcolor{black}{
It is interesting to observe that these methods achieve different performance on the two datasets. For example, the method of~\cite{yoo2019pseudoedgenet} achieves unsatisfactory performance on MoNuSeg, which it works well on CPM.
The reason may be that the MoNuSeg dataset is more challenging than CPM, in which the nuclei are more densely distributed and the background is more complex.
The method of~\cite{yoo2019pseudoedgenet} mainly relies on the edge information of nuclei extracted from the Sobel filter, which is hard to split the touching nuclei.
In contrast, the methods of~\cite{Qu2020,chamanzar2020weakly,lee2020scribble2label} are based on the Voronoi diagram that splits the point annotation into multiple convex polygons, which is intrinsically able to handle the touching nuclei.
}

\textcolor{black}{We believe that the superior performance owes to the proposed co-training strategy and the colorization proxy task. Compared with~\cite{xie2020instance,lee2020scribble2label}, which perform self-training with nuclei size and quantity, our method utilizes the co-training strategy in which the two models mutually facilitate each other in a bootstrapping way, so as to gradually improve the segmentation results. Compared with~\cite{tian2020weakly,Qu2020,chamanzar2020weakly}, which rely heavily on the derived ambiguous labels (i.e., the Gaussian point masks, the Voronoi label, and the cluster label), our method further adopts the auxiliary task of colorization to extract the unique color and texture information of H\&E stained images, thereby acting as effective supervision information of shape and boundary of nuclei, guiding the segmentation task. In addition, using the gray-scale images obtained by extracting the hematoxylin component as inputs alleviates the requirements of additional boundary supervision, however our method can still accurately identify the nucleus boundary without introducing the contour map for edge refinement~\cite{yoo2019pseudoedgenet}. 
{Using the H-component as input could potentially reduce the affect of color variations~\cite{zhao2020triple}, we therefore did not perform additional staining normalization/augmentation in our experiments.}
In addition, our method greatly reduces the annotation efforts as shown in a study~\cite{Qu2020} that point annotation reduces 88\% annotation workload compared to full pixel-wise annotation.}

\begin{figure}[!t]
    \centering
    \includegraphics[width=0.48\textwidth]{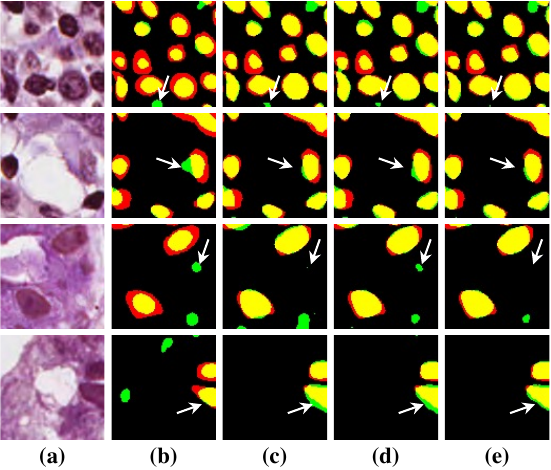}
    \caption{Nuclei segmentation results of several variants of the proposed method. (a) Original H\&E images; the ablation settings of (b), (c), (d), and (e)correspond to Models A, B, C, and D in Table~\ref{tab_ablation_study}, respectively. Yellow, red, and green denote true positive, false positive, and false negative, respectively. White arrows indicate the significant differences.}
    \label{fig_visual}
\end{figure}
\begin{figure}[htbp]
	\centering
	\subfigure[]{
		\includegraphics[width=0.10\textwidth]{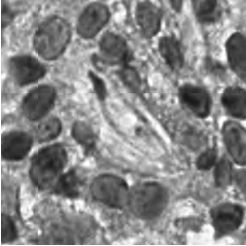}
	}
	\subfigure[]{
		\includegraphics[width=0.10\textwidth]{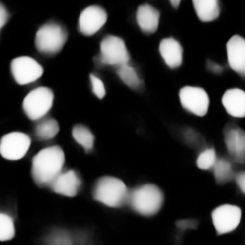}
	}
	\subfigure[]{
		\includegraphics[width=0.10\textwidth]{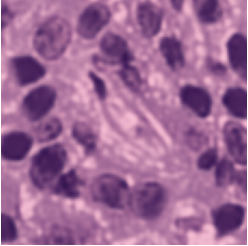}
	}
	\subfigure[]{
		\includegraphics[width=0.10\textwidth]{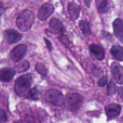}
	}
	\caption{Image colorization.
		(a) The input of the model with H-component; (b) The predicted segmentation result; (c) The predicted colorization result; (d) The H\&E stained image.}
	\label{fig.5}
\end{figure}
\begin{table}[t]
	\centering
	\caption{Ablation Study of the proposed method. The \checkmark indicates that the corresponding loss (i.e., Voronoi loss $L_{vor}$, cluster loss $L_{clu}$, co-training loss $L_{cot}$, and colorization loss $L_{color}$) is included in the total loss.}
	\setlength{\tabcolsep}{1pt}{
		\begin{tabular}{ccccccccc}
			\toprule
			Model & $L_{vor}$ & $L_{clu}$ & $L_{cot}$  & $L_{color}$ & Acc (\%)  & F1 (\%)  &  $\mathrm{Dice}_{obj}$ (\%) & AJI (\%) \\
			\midrule
			A & \checkmark &        &        &        & 87.91 & 62.86 & 60.20 & 43.01 \\
			B & \checkmark &\checkmark &        &        & 90.60 & 77.39 & 72.48 & 51.77  \\
			C & \checkmark &\checkmark &\checkmark &        & 91.26 & 77.64 & 72.87 & 52.96 \\
			D & \checkmark &\checkmark &\checkmark &\checkmark & \textbf{91.44} & \textbf{77.64} & \textbf{74.41} & \textbf{56.20}  \\
			\bottomrule
		\end{tabular}}%
	\label{tab_ablation_study}%
\end{table}%

\subsubsection{Ablation Studies}
\textcolor{black}{
To investigate the impact of the proposed co-training and colorization methods, we validate the effectiveness of each module on the MoNuSeg dataset. Considering that Voronoi labels provide more shape supervision information than point annotation, we regard the model that jointly uses the point annotation and the Voronoi label as the baseline, and mainly evaluate the improvement of each module upon this baseline in Table~\ref{tab_ablation_study}.
\textit{Model A} utilizes only point annotation and Voronoi labels with cross-entropy loss. For a fair comparison with co-training, we employ two separate models which have the same architectures for co-training and average the predictions of the two models in inference. \textit{Model B} further adopts the cluster label. It can be seen that the cluster loss is beneficial in segmenting overlapping nuclei, improving the $\mathrm{Dice}_{obj}$ from 60.20\% to 72.48\% and the AJI from 43.01\% to 51.77\%, which indicates that simply using the point annotation and the derived Voronoi labels cannot provide sufficient supervision for the nuclei segmentation due to the lack of boundary and area information. However, the cluster labels generated by the $k$-means algorithm cannot separate close nuclei which would introduce distraction to the training process, as shown in Fig.~\ref{fig_label_clu}. To address this challenge, \textit{Model C} utilizes the co-training strategy that the two models facilitate each other in a bootstrapping way to eliminate the distraction brought by the cluster labels. As expected, co-training brings a consistent improvement in all metrics, including an 1.19\% increase to AJI. We believe that co-training compensates for the loss of supervision information in the uncertain regions in the cluster label, and the probability map with EMA by another model provides more precise and robust supervision in the nuclei boundary. Visualization of the segmentation results can be found in Fig.~\ref{fig_visual}, and it can be seen that with the co-training strategy, the model can achieve more accurate results, especially in the nuclei boundary area. \textit{Model D} involves colorization as a proxy task to implicitly learn to be self-aware of the nuclei boundary. Instead of directly coloring the H-component (Fig.~\ref{fig.5}(a)) to the H\&E map (Fig.~\ref{fig.5}(d)), we use the segmentation probability map (Fig.~\ref{fig.5}(b)) as the input of the colorization network, which can help us to boost the segmentation accuracy. The experimental result shows that integrating the colorization tasks could not only improve the $\mathrm{Dice}_{obj}$ by 1.54\% and AJI by 3.24\%, but also promote Acc by 0.18\% without dropping F1 score, proving that the colorization task has a significant guiding effect on nuclei segmentation. In Fig.~\ref{fig.5}(c), the improvement of colorization in the nuclei boundary area can also be observed. In general, the four modules used in our method have complementary advantages. By minimizing the weighted sum of the four losses,  the proposed framework can distinguish between nuclear and non-nuclear to the greatest extent.
}

\subsubsection{Comparison with Fully-Supervised Methods}
\textcolor{black}{We compare our weakly-supervised method with five fully-supervised methods that are trained with the completely-annotated nuclei masks, such as ResUNet~\cite{Qu2020}, U-Net~\cite{ronneberger2015u}, FCN~\cite{long2015fully}, PSPNet~\cite{zhao2017pyramid} and DeepLab~\cite{2018DeepLab}. The segmentation results are shown in Table~\ref{tab_model_fully}. It can be seen that ResUNet achieves the best performance with 80.23\% in F1 score and 58.93\% in AJI\footnote{Note that for a fair comparison, we use the binary labels instead of instance labels, and discard the tricks to separate the touching nuclei (e.g., test time augmentation (TTA), model ensemble, multi-branch framework, heavy-weight backbone, post-processing steps).
The fully supervised ResUNet achieves 58.93\% in AJI, outperforming the baseline CNN3~\cite{kumar2017dataset} and DIST~\cite{naylor2018segmentation} by a margin of 8.10\% and 2.95\%, respectively.}. And with only point annotation, our method achieves competitive performance with 74.41\% in $\mathrm{Dice}_{obj}$ and 56.20\% in AJI, outperforming three out of the five fully-supervised methods, which verifies the effectiveness of our method.
We also report the performance of several state-of-the-art methods~\cite{kumar2017dataset,raza2019micro,graham2019hover,chen2020boundary,schmidt2018cell,wang2022multi,he2021cdnet,zhou2019cia} on MoNuSeg and CPM dataset. It can be seen that there still exists non-negligible performance gaps between the fully-supervised and weakly-supervised methods.
}

\begin{table}[t]
	\centering
	\caption{Comparison of the proposed weakly-supervised method with fully-supervised methods.}
	\resizebox{0.48\textwidth}{!}{
	\begin{tabular}{l|cc|cc}
		\toprule
            \multirow{2}{*}{Methods} & \multicolumn{2}{c|}{MoNuSeg} & \multicolumn{2}{c}{CPM} \\
            \cline{2-5}
		& F1 (\%)  & AJI (\%)  & F1 (\%)  & AJI (\%)  \\
		\midrule
		FCN~\cite{long2015fully} & 73.92 & 54.58 & 80.39 & 47.88     \\   
		PSP~\cite{zhao2017pyramid} & 73.64 & 53.54 & 79.91 & 47.89     \\
		DeepLab~\cite{2018DeepLab} & 74.45 & 54.39 & 80.28 & 48.61     \\
		U-Net~\cite{ronneberger2015u} & 75.75 & 57.38 & 81.36 & 55.16     \\  
		ResUNet~\cite{Qu2020} & 80.23 & 58.93 & 81.50 & 55.63     \\
		\midrule
            CNN3~\cite{kumar2017dataset} & 76.20    & 50.83 & - & - \\
            DIST~\cite{naylor2018segmentation} & 78.63 & 55.98 & - & - \\
            Micro-Net~\cite{raza2019micro}  & 81.90	& 60.90	& 85.70	& 66.80 \\
            Hover-Net~\cite{graham2019hover}   & 82.60 & 61.80	& 86.90	& 70.50 \\
            BRP-Net~\cite{chen2020boundary} & 82.10	& 64.22	& 87.70	& 73.10 \\
            StarDist~\cite{schmidt2018cell} & 82.20	& 61.90	& -	& - \\
            RCSAU-Net~\cite{wang2022multi} & 82.00	& 61.90	& 88.00	& 72.70 \\
            CDNet~\cite{he2021cdnet} & 83.16	& 63.31 	& 88.0	& 73.26 \\
            \midrule
            \multicolumn{5}{c}{Top 5 solutions in MoNuSeg challenge.} \\
            \midrule
            CIA-Net~\cite{zhou2019cia}  & - & 69.07 & 84.16 & 66.48 \\
            BUPT.J.LI   & - & 68.68	& -	& - \\
            pku.hzq	& -	& 68.52	& -	& - \\
            Yunzhi	& -	& 67.88	& -	& - \\
            Navid Alemi	& -	& 67.79	& -	& - \\
		\midrule
		\textbf{Ours}   & 77.64  & 56.20 & 79.72  & 51.32 \\
		\bottomrule
    \end{tabular}}
	\label{tab_model_fully}%
\end{table}%

\begin{table}[htbp]
	\centering
	\caption{Effect of different training-data-split strategies.}
    \setlength{\tabcolsep}{2.5mm}{
		\begin{tabular}{c|cccc}
			\toprule
			Overlap  & Acc (\%)  & F1 (\%)  & $\mathrm{Dice}_{obj}$ (\%)  & AJI (\%)\\
			\midrule
		    0         & \textbf{91.44}  & 77.64   & \textbf{74.41}  & \textbf{56.20}\\
		    20\%      & 91.31     & 77.32         & 73.59     & 54.69\\
		    40\%      & 91.42     & 77.28         & 73.51     & 54.65\\
		    60\%      & 91.25     & 77.01         & 72.57     & 53.05\\
		    80\%      & 91.41     & \textbf{77.84}         & 73.61     & 54.13\\
		    100\%     & 91.04    & 77.54    & 73.17   & 53.02\\

			\bottomrule
	    \end{tabular}}
	\label{tab_dff_data}%
\end{table}%
\subsection{Analyses of Training-Set Split Strategies}
To evaluate the effect of the data split strategy of the two models, we conduct a series of experiments with different overlaps of the two subsets, and the results are shown in Table~\ref{tab_dff_data}. We investigate three data-split strategies: completely non-overlapping, partially overlapping, and completely overlapping. 
Note, the size of subsets without overlap (i.e., overlap ratio of 0\%) is a half of the subsets with complete overlap (i.e., overlap ratio of 100\%). It can be seen that the best result is achieved when the two subsets do not overlap at all, and a partial overlap outperforms the complete overlap. We believe that for two identical models that are co-trained, the same data makes the image features learned by the two models the same, resulting in no mutual supervision effect. However with non-overlapping training sets, the two models can obtain additional complementary information by learning from different images, and effective mutual supervision benefits each other. The greater the difference between the two subsets, the stronger the effect of mutual supervision.

\begin{figure}[!t]
	\centering
	\subfigure[]{
		\includegraphics[width=0.2\textwidth]{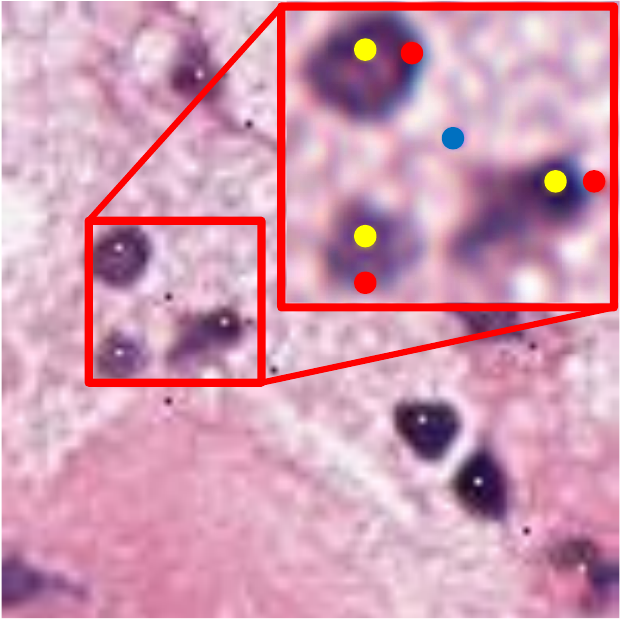}
	}
	\subfigure[]{
		\includegraphics[width=0.2\textwidth]{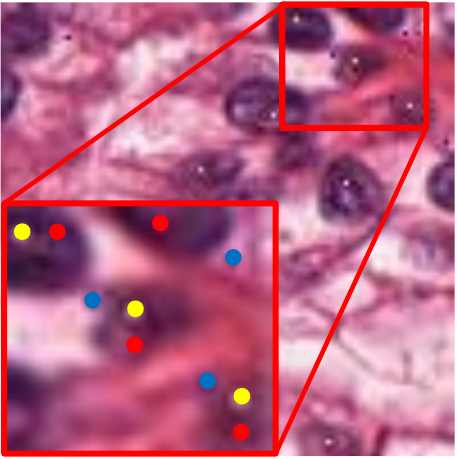}
	}
	\caption{The visualization of shifted annotation points. (a) and (b) are patches from two images. Yellow, red and blue points are the central points, points offset by four pixels and eight pixels, respectively.}
	\label{fig_shift}
\end{figure}

\begin{figure}[!t]
	\centering
	\subfigure[]{
		\includegraphics[width=0.22\textwidth]{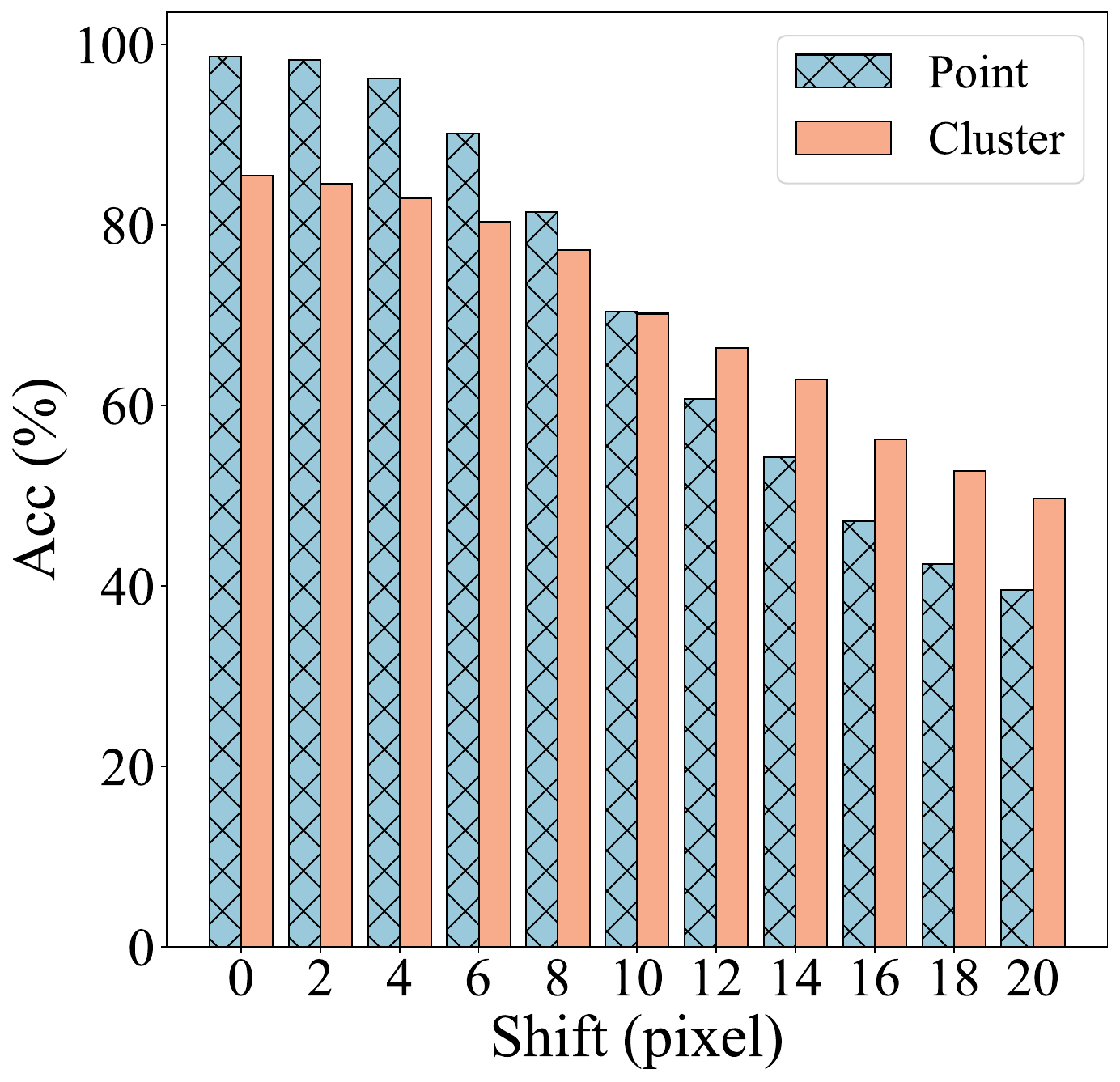}
	}
	\subfigure[]{
		\includegraphics[width=0.22\textwidth]{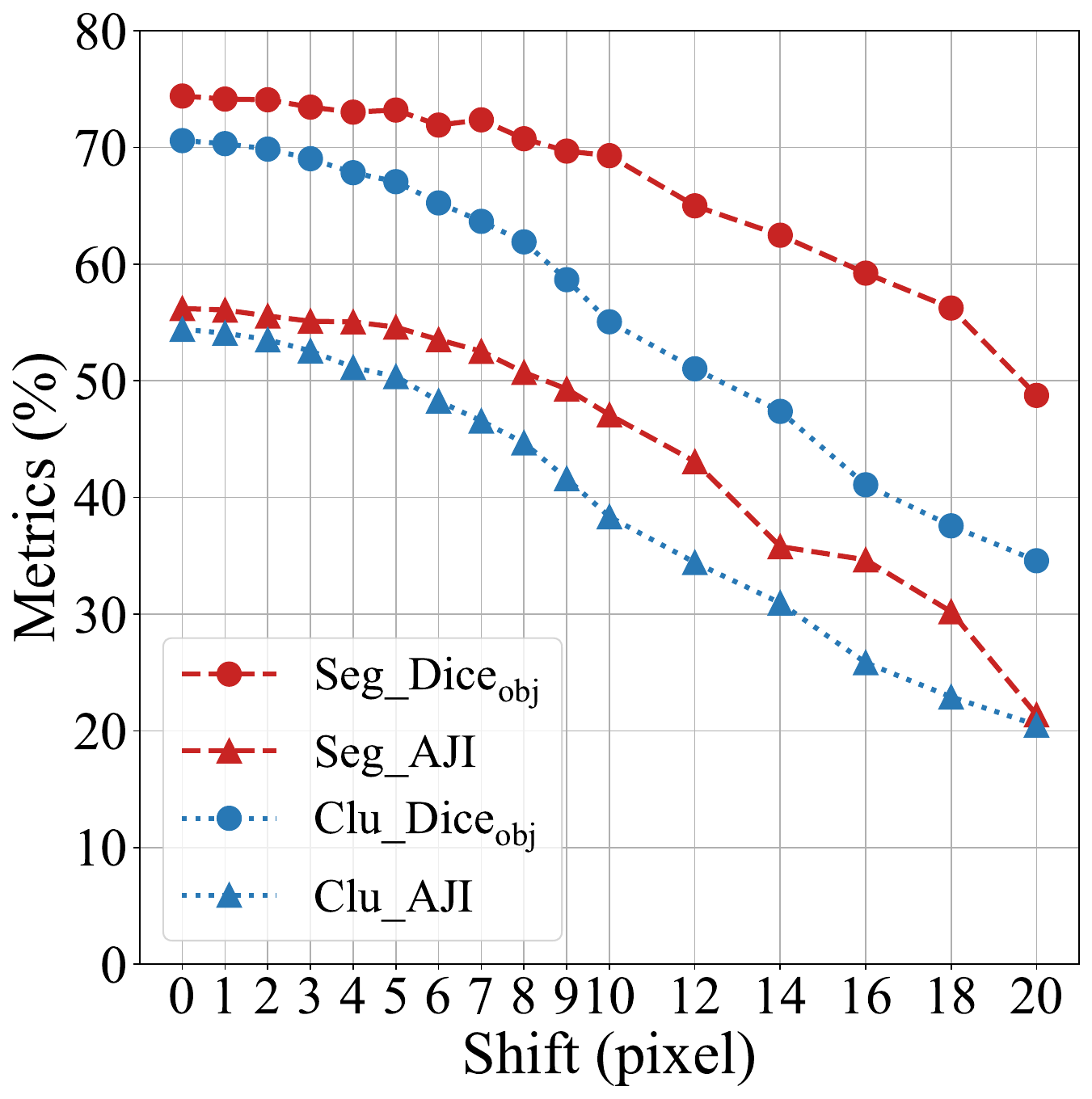}
	}
	\caption{Quantitative results with different shifts in point annotation. (a) Accuracy of point annotations and cluster labels; (b) Object-level metrics of segmentation results (Seg) and cluster results (Clu).}
	\label{fig_shift_results}
\end{figure}

\begin{figure}[htb]
    \centering
    \color{black}
    \includegraphics[width=0.4\textwidth]{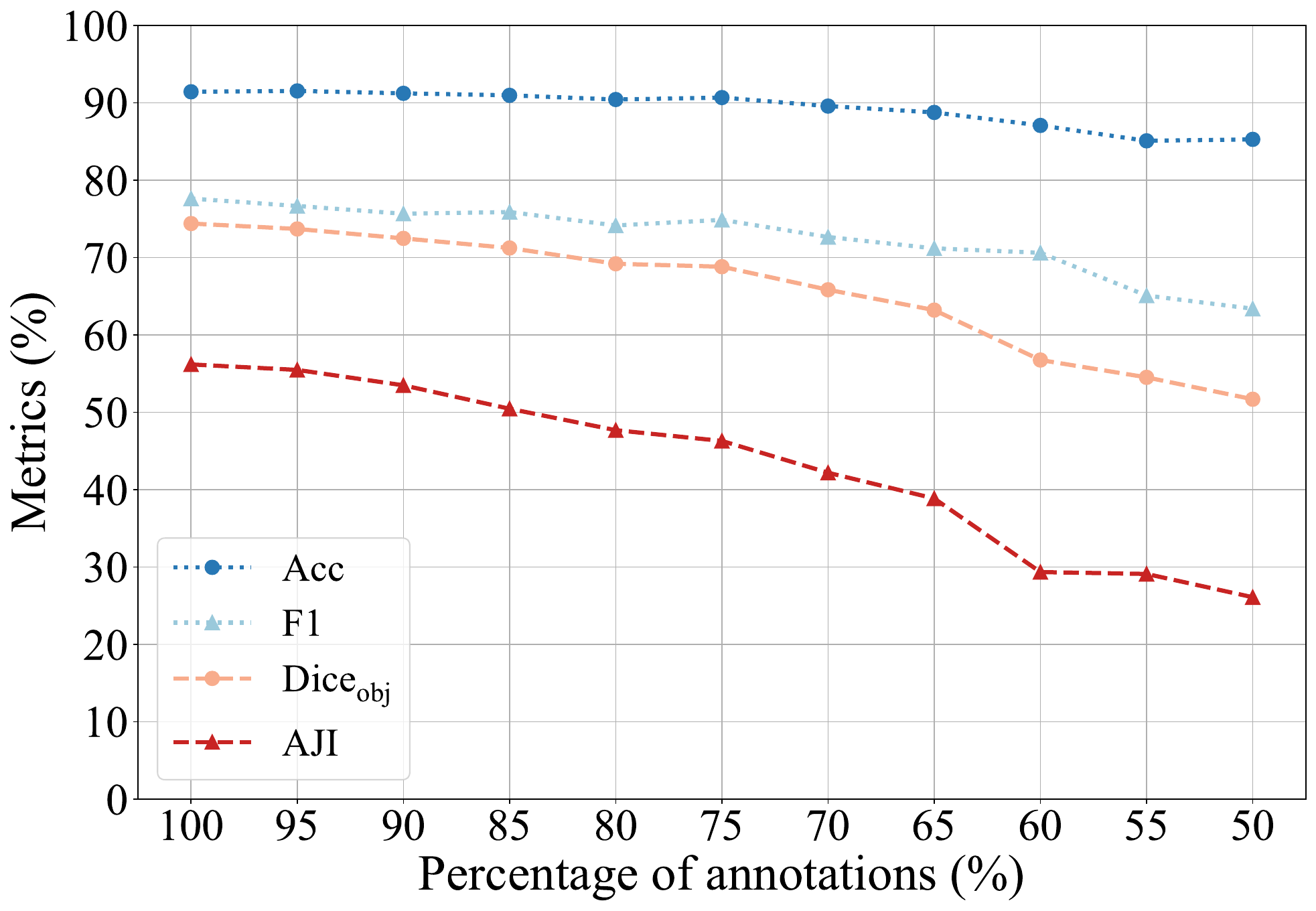}
    \caption{Quantitative results with different ratio in point annotation.}
    \label{fig_completeness}
\end{figure}

\begin{figure*}[!t]
    \centering
    \includegraphics[width=0.98\textwidth]{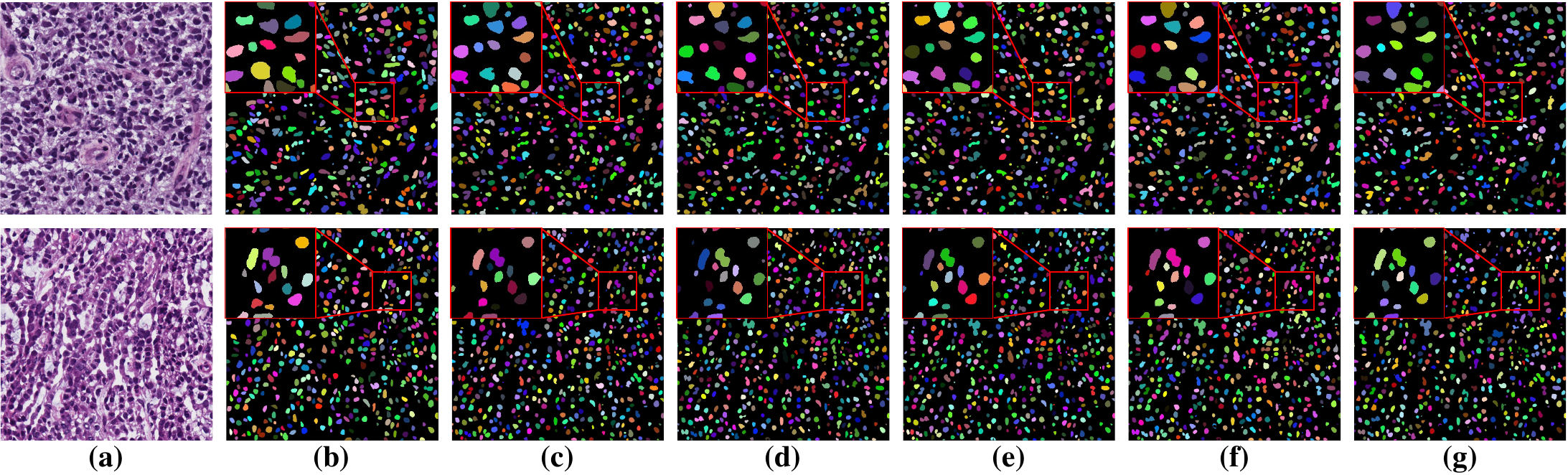}
    \caption{Instance segmentation results with different shifts of point annotations. (a) H\&E stained image, (b) ground-truth mask, (c)-(g) are results using the point annotations with shift of 2, 4, 6, 8 and 10 pixels, respectively.}
    \label{fig_ins_shift}
\end{figure*}

\subsection{Impact of the Perturbation in Point Annotation}
\label{sec:impact_perturbation}
In clinical practice, due to the time constraint, pathologists cannot exactly put the annotation point at the center of a nucleus. We carry out experiments to investigate the impact of point perturbation to segmentation performance.
To simulate the actual annotations, we perform a uniform random shift within different ranges of the generated point annotations. Several examples are shown in Fig.~\ref{fig_shift} illustrating that the small shift makes the points not too far from the center, but there are still some cases that the points are close to the boundary or even outside the nuclei. The number of points falling outside of the nuclei increases as the shift increases. Fig.~\ref{fig_shift_results}(a) gives a quantitative illustration that as the shift increases from 0 to 20 pixels, the ratio of point annotations being within the nuclei decreases from 98.66\% to 39.54\%. Obviously, the offset of the points reduces the quality of the coarse labels, especially the accuracy of the nuclei in cluster labels is reduced from 85.46\% to 49.72\%.
We train the models with the point annotations obtained by different shifts and calculate two object-level metrics. 
As illustrated by the changing trends of $\mathrm{Dice}_{obj}$ and AJI versus the shifts in Fig.~\ref{fig_shift_results}(b), 
the segmentation accuracy of the proposed method degrades more gradually than that of the cluster results. The four-pixel shift only result in 1.38\% drop in $\mathrm{Dice}_{obj}$ and 1.15\% drop in AJI to the segmentation results of the proposed method, but 2.75\% drop in $\mathrm{Dice}_{obj}$ and 3.27\% drop in AJI to the cluster results. Even though the eight-pixel shift makes about one fifth point annotations not in the nuclei, the segmentation performance can still reach 70.75\% in $\mathrm{Dice}_{obj}$ and 50.72\% in AJI, respectively, which verifies the robustness of our method to point annotation offset. We believe that the robustness owes to the label propagation which refines the coarse labels to compensate for the impact of the offset. 
\textcolor{black}{
In clinical practice, professional pathologists will ensure the quality of point annotations, thus avoiding performance degradation caused by excessive offset.
}
Some nuclei instance segmentation results with different shifts of point annotation are presented in Fig.~\ref{fig_ins_shift}. The robust instance segmentation performance lays the foundation for further counting and morphological feature extraction of nuclei on pathology images.

\subsection{Impact of the Completeness in Point Annotation}
\textcolor{black}{
As an important component of our method, Voronoi diagram is based on two assumption: 1) the point annotation is at the center of the nuclei; 2) all the nuclei are labeled. 
The impact of the first assumption has been studied in Sec.~\ref{sec:impact_perturbation}. 
This section investigates the robustness of the proposed model to the completeness of the point annotation.
As shown in Fig.~\ref{fig_completeness}, to verify what level of completeness is required to train this model without a significant loss of performance, we gradually reduce the percentage of point annotations for training, and then test and calculate the segmentation metrics. When only 1\% of the point annotations were missing, the segmentation metrics are only slightly reduced, as $\mathrm{Dice}_{obj}$ decreases from 74.41\% to 74.18\% and AJI decreases from 56.20\% to 56.15\%. As the percentage of annotations decreases, segmentation metrics reduce. However, our method can still maintain good performance when 15\% of the annotations are missing. The missing of too many annotations will greatly reduce the accuracy of the generated coarse labels, and then lead to the reduction of segmentation performance.}

\section{Discussion}
\textcolor{black}{
This section discusses the contributions of the proposed method, compared with the most relevant works. One of the main contributions of the proposed method is that it achieves comparable performance to the fully-supervised methods, while significantly reducing the annotation efforts (e.g., reducing the annotation time by 88\%~\cite{Qu2020}). 
This paper provides a promising solution for weakly-supervised nuclei segmentation with point label and could facilitate the community to further advance the research in the field of nuclei analysis in pathology images.}

Existing methods typically adopted the conventional methods based on the shape and contrast prior of nuclei to transform the point annotation into the coarse pixel-level annotations to facilitate the training of deep learning models. However, the performance could suffer from the incomplete and inaccurate coarse labels.
To address this challenge, several studies~\cite{qu2019weakly,yoo2019pseudoedgenet} proposed to inject additional supervision by magnifying the local contrast changes to locate the boundary of nuclei, and the others~\cite{chamanzar2020weakly,Qu2020} attempted to carefully design a multi-stage training strategy to boost the segmentation performance. However, the existing methods have two limitations: 1) the additional supervision cannot guarantee the correctness and may introduce more distractions; 2) the multi-stage training strategy may cause error accumulation. 

Compared to the existing methods, the advantages of this work are as follows. First, we design a co-training strategy that enables the two collaborative networks to transfer knowledge between each other, so as to compensate for the missing supervision in the coarse labels. The pseudo label produced by the segmentation network can be gradually refined during the training process, avoiding the misleading of the additional supervision. To the best of our knowledge, the work of Zhao et al.~\cite{zhao2020institute} is the most relevant to our co-training strategy. The authors proposed a divergence loss to encourage the diversity between the two models, which however may consequently hurting the performance of each model. On the contrary, we propose a pseudo label generation method based on EMA to periodically average the predictions to supervise the other model, which alleviates the problem of self-deception and produces more robust pseudo labels. Second, we customize a self-supervised learning method for nuclei segmentation in pathology images.
Different from the related methods~\cite{koohbanani2021self,li2021dual,srinidhi2022self} that simply applied the SSL techniques from natural images, our method appreciates the special characteristics of the H\&E staining method for pathology images, involving the prior knowledge of nuclei. 
The most relevant work of our nuclei-aware colorization method is Yang et al.~\cite{yang2021self}, which designed two pretext tasks transforming between the H-component and E-component. On the contrary, our method employs a novel framework, i.e., SC-Net, which sequentially combines the segmentation network with the colorization network, enabling the colorization network to benefit the segmentation task during the whole training process in an end-to-end manner. 
Third, instead of manually scheduling the training process in multiple stages, we introduce cumulative learning to integrate different modules in this pipeline. 
In this way, the different modules can be integrated into a single shot training procedure.

\textcolor{black}{
There are several potential use-cases where point annotation is essential. To name a few, for a locally collected dataset, point annotation is a much more convenient way for pathologists to train the nuclei segmentation model. Second, our method based on point annotation shows competitive performance, which could be used to help the pathologist revised the annotation or label the data in a semi-automatic manner. Third, in Sec.~\ref{sec:impact_perturbation}, the experiment shows that our method is robust to the perturbation in point annotation, which could facilitate the annotation procedure for junior pathologists. 
}

\section{Conclusion and future work}
In this paper, we studied the problem of weakly-supervised nuclei segmentation with only point annotation, addressing the concern that manual annotation of nuclei/cell in pathology images is time-consuming and labor-intensive. To this end, we presented an annotation-efficient framework based on the idea of label propagation to fully incorporate the prior knowledge and explore the potential supervision information of point annotation. The proposed method consists of three parts: 1) coarse pixel-level label generation from the point labels; 2) a deep co-training strategy with EMA to smoothly transfer the knowledge between the two collaborative networks; 3) self-supervised learning tailored for pathology images with a novel auxiliary network to guide and verify the segmentation network. We validated our method on two public datasets, i.e., MoNuSeg and CPM, and both quantitative and qualitative evaluations demonstrated the superiority of our method to the state-of-the-art methods and the effectiveness of each module.

There are several limitations in this study. First, the point annotations of this study are supposed to be complete, which means it requires all the nuclei to be labeled, otherwise the generated coarse labels would collapse. Second, the co-training method increases the demand for memory and computational power. We will try to enhance the method with greater computational efficiency without compromising the segmentation accuracy.

Besides that, we have identified a few further directions for future studies. First, we plan to explore the abundant unlabeled data in the semi-supervised learning scenario. Second, we will further aim at the integration of different label granularities to fully leverage as much as possible from affordable annotations. Third, the touched and overlapped nuclei in pathology images are challenging to segment at the instance level, leaving for future work to further develop methods for nuclei instance segmentation. And last but not least,  we will try to adapt the proposed method for interactive annotation to further reduce the annotation burden of pathologists, promoting the research of automatic pathology image analysis.

\bibliographystyle{IEEEtran.bst}
\bibliography{refs}

\end{document}